\documentclass[bm]{epl}

\usepackage{graphicx}
\usepackage{bm}

\title{Dynamic surface decoupling in a sheared polymer melt}

\shorttitle{Surface decoupling in a sheared polymer melt}

\author{Xin Zhou \and Denis Andrienko \and Luigi Delle Site \and Kurt Kremer}
\shortauthor{Xin Zhou \etal}

 \institute{Max-Planck-Institut f\"{u}r Polymerforschung -
            Ackermannweg 10, 55128 Mainz, Germany }


\begin{document}

\maketitle

\begin{abstract}
  We propose that several mechanisms contribute to friction in a
  polymer melt adsorbed at a structured surface. The first one is the
  well known disentanglement of bulk polymer chains from the surface
  layer. However, if the surface is ideal at the atomic scale, the
  adsorbed parts of polymer chains can move along the equipotential lines of
  the surface potential. This gives rise to a strong slippage of
  the melt. For high shear rates chains partially desorb.
  However, the friction force on adsorbed chains increases, resulting
  in quasi-stick boundary conditions. We propose that the
  adsorbed layers can be efficiently used to adjust the
  friction force between the polymer melt and the surface.
\end{abstract}

\section{Introduction}

Frictional forces between solid substrates can be dramatically reduced
by coating them with thin polymer layers, which are able to sustain
large normal loads while remaining fluid~\cite{klein1996}. This effect
has obvious practical applications, ranging from biolubrication to
hard disk drives~\cite{jin1993}.
The molecular mechanism of friction, however, is still poorly
understood~\cite{grest1999}. It is believed that the
striking reduction of shear forces in systems with end-adsorbed and
grafted polymers immersed in liquids is due to limited mutual
interpenetration of the brushes in good solvent, even under
compression, and a fluid interfacial layer where most of the shear
occurs. In polymer melts the situation is very different, because of
the dominance of entanglement effects (if the chains are long enough),
associated high viscosities, and screening of the excluded volume
interactions~\cite{yamamoto2004,everaers2004}. In spite of these
differences, the key parameter affecting the lubrication and shear
forces is still the degree of interdigitation between the moving
surface layer and the bulk polymer system~\cite{leger1999}.
Regarding the structure of the adsorbed polymer layers two systems
were thoroughly examined, both theoretically and experimentally:
(i) Polymer brushes with the chains tethered by one end to a solid
surface. At high enough coverage, the chains stretch away from the
surface and form a brush-like structure. 
%
%
(ii) Melt layers which are formed when a strongly attractive surface
is exposed to a polymer melt and a certain amount of polymer becomes
permanently bound to the surface. The resulting layer is made of
loops, with a large polydispersity of loop sizes, reflecting the
statistics of the chains in the melt~\cite{guiselin1992,smith2005}.
%
As to the rheological properties of brushes, it is now well
accepted that the polymer anchoring on solid surfaces plays a key
role on the flow behavior of polymer melts, in particular on the
appearance of flow with {\em slip} at the wall~\cite{brochard1992,ajdari1994}.  
The onset of wall
slip is related to the strength of the interactions between the
solid surface and the melt: if there is a finite slip velocity
$v_s$ at the interface, the shear stress at the solid surface can
be evaluated as $\tau_{xz} = \beta v_s$, where $\beta$ is the
friction coefficient between the fluid molecules in contact with
the surface and the solid surface. On the other hand, $\tau_{xz} =
\eta(dv/dz)_{z=0}$, where $\eta$ is the melt viscosity.
Introducing the extrapolation length (slip length) $b$ of the
velocity profile to zero ($b=v_s / (dv/dz)_{z=0}$), one obtains
$\beta=\eta/b$. Thus, determination of $b$ will yield $\beta$, the
friction coefficient between the surface and the fluid, which is
directly related to the molecular interactions between the fluid
and the solid surface~\cite{leger1999}.

The appearance of wall slip in polymer melts can be expected from the analogous 
phenomenon in simple liquids~\cite{thompson1997,barrat1999},
which still pose unresolved challenges in understanding~\cite{cottinbizonne2003}.
Typical model systems assume that chain ends (or parts of the chain)
are fixed at the surface\cite{ajdari1994,smith2005,brochardwyart1994}. However, reasonably weak coupling might result in partial or complete detachment of the chains. This certainly will be the case for most adsorbing chain ends or fragments (i.~e. for block copolymers). In addition, the attached parts of the chains can
travel along the equipotential lines of the surface or along minimal
activation paths. In this work we thus focus on the situation where
the surface bonding is strong but reversible. As an example, we
consider a bisphenol-A polycarbonate (BPA-PC) melt sheared over a
(111) nickel surface. The inner parts of the polymer chains are
repelled from the surface and only one of the comonomeric groups
(phenol) is attracted to it.  Internal phenylenes are, however,
sterically hindered to adsorb by the rest of the groups.  Contrary,
chain ends can easily adsorb to the surface; a large adsorption energy
(about $20 kT$ at the processing temperature of $570\,\rm K$) prevents
the adsorbed chains from a spontaneous desorption. A closer look at
the structure and packing of short but experimentally relevant polymer
chains at the walls shows that there are two distinct (overlapping)
layers, formed by the single- and two-end adsorbed chains. The layer
composed of the two-end adsorbed chains is made of highly entangled
loops, with the configurations reflecting the statistics of the chains
in the melt. The single-end adsorbed chains are stretched and form a
brush-like layer, which is strongly interdigitated with the bulk of
the melt~\cite{abrams2003,dellesite2002,abrams2003b,dellesite2004}.
%
The complex structure of the surface layer enriches the behavior
of the sheared melt: in addition to large slip length and
decoupling of the surface layer from the bulk, we also observe the
transient regime, when the two-end attached chains are practically
unaffected by shear, but the single-end attached chains are
stretched and form a (lubricating) layer between the
two-end-adsorbed and bulk chains.  For even higher shear rates we
observe shear-induced desorption of the single-end adsorbed
chains, which affects both the width of the adsorbed layer and the
value of the slip length.

\section{Model and interaction potentials}
\begin{figure}
  \onefigure[height=2cm]{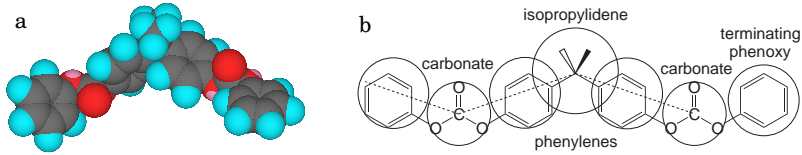}
\caption{
  (a) chemical structure and (b) coarse-graining scheme of
  bisphenol-{\em A} polycarbonate. Corresponding comonomeric groups
  (isopropylidene, carbonate, and two phenylenes) are replaced by four
  beads. Only the chain end and one repeat unit are shown.  }
  \label{fig:bpa-pc}
\end{figure}

The coarse-grained model is discussed in previous
works~\cite{abrams2003,abrams2003b}. Briefly, each repeat unit
(carbonate, isopropylidene, and two phenylenes) of a BPA-PC chain is
replaced by four different beads, as shown in Fig.~\ref{fig:bpa-pc}.
All beads interact with each other via shifted repulsive Lennard-Jones
potential. The sizes of the beads are such that the volume of the
repeat unit, accounting for overlaps, is equal to the van der Waals
volume per repeat unit computed from the equation-of-state
analysis~\cite{tschop1998}. Intramolecular bond-angle and harmonic
spring potentials reflect the Boltzmann distribution of the
corresponding variables for a single chain in vacuum.


The bead-wall interaction potential reflect the structure of the surface. 
%
%
Expanding this potential in two-dimensional reciprocal lattice
space and performing the summation over the reciprocal vectors of
the same length we obtain
%
\begin{eqnarray}
U(x,y,z) = \sum_{i} U_{i}(z) \, f_{i}(x,y),
\label{eq:potential}
\end{eqnarray}
where $i=0,1,2,...$ corresponds to the reciprocal vectors of different
lengths,
$f_{i}(x,y)$ are linear combinations of trigonometric functions.
Expansion~(\ref{eq:potential}) can be used for a surface of any
symmetry. To proceed further, we make use of the fact that the
(111) surface of the nickel fcc lattice has the $C_6$ symmetry.
Restricting ourselves to the first three terms in
(\ref{eq:potential}) we obtain $f_{0} = 1$, $f_{1}=\cos({\bar
x}-{\bar y})+\cos({\bar x}+{\bar y})+\cos 2{\bar y}$, and
$f_{2}=\cos({\bar x}-3{\bar y})+\cos({\bar x} + 3{\bar y})+ \cos 2
{\bar x}$, where $({\bar x},{\bar y}) = {2 \pi \over a}\,(x,{y
\over {\sqrt 3}})$.
%
%
The $z$-dependent prefactors $U_i(z)$ are obtained from the {\em ab~initio} calculations~\cite{dellesite2003}. These calculations show that
the adsorption energies of both carbonate and isopropylidene are
rather small, much smaller than the characteristic thermal energy
in a melt at processing temperature; below $3.2\,{\rm \AA}$, they
experience strong repulsion from the nickel surface. Contrary, the
phenolic group is attracted by the surface, with adsorption energy
about $1\, {\rm eV}$ at an optimal distance $z_0=2\,{\rm \AA}$;
the adsorption is short ranged and decays below $0.03\,{\rm eV}$
at a distance beyond $z_c=3\,{\rm \AA}$.  Because of that the
internal phenylenes are sterically hindered by the neighboring
comonomeric groups to adsorb. We assume that 
they interact with the walls via a 10-4 repulsive potential.
The chain ends are not
hindered by their nearest neighbors. The results of the {\em
ab~initio} calculations suggest a simple piecewise function for
the attractive surface-phenoxy bead interaction potential at a
bridge site~\cite{abrams2003b}
\begin{eqnarray}
\nonumber
U_{0} &=& \left\{
\begin{array}{lll}
 \frac{5}{3} {\epsilon}_{r}
\left[ \frac{2}{5}
\left( \frac{z_0}{z} \right)^{10} -
\left( \frac{z_0}{z} \right)^4 + \frac{3}{5} \right] - \epsilon_0,
&  z < z_{0}  \\
 \frac{\epsilon_0}{2} \left[
\cos \left(\pi \frac{z_c - z}{z_c-z_0} \right)-1 \right],
&  z_0 \le z < z_c
\end{array}
\right.
\end{eqnarray}

\begin{figure}
\twoimages[height=4.5cm]{fit_cpmd.eps}{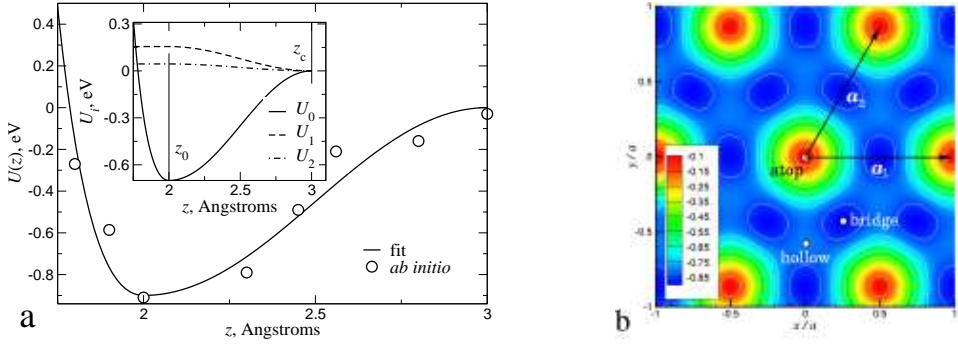}
\caption{
  (a) Wall-phenoxy bead interaction potential taken at a bridge site. Open
  circles: results of {\em ab~initio} calculations. Solid line: fit to
  Eqn.~\ref{eq:potential}. Inset shows the pre-factors $U_{0,1,2}(z)$.
  (b) Contour plot of the attractive potential between the terminating
  phenoxy and the wall at a distance $z_0 =\,{\rm 2\AA}$.  Three
  values matched the {\em ab~initio} data: $U_{\rm atop} = -0.1\,{\rm
    eV}$, $U_{\rm bridge} = -0.9\,{\rm eV}$, $U_{\rm hollow} = -0.8\,
  {\rm eV}$.  }
\label{fig:fitcpmd}
\end{figure}

{\em Ab~initio} results also suggest that the {\em functional
form} of the $z$-dependent part of the potential is practically
independent from the $xy$ position of the bead, only the well
depth ($\epsilon_0$) of the potential changes. At an optimal
distance $z = 2\,{\rm \AA}$ the adsorption energies are: $U_{\rm
atop} = -0.1\,\rm eV$, $U_{\rm bridge} = -0.9\,\rm eV$, $U_{\rm
hollow} = -0.8\, \rm eV$~\cite{dellesite2003}.  Adopting these
values and assuming the same functional form for the attractive
part of $U_{0,1,2}(z)$ we obtain
\begin{equation}
\nonumber
U_{1,2} = \left\{
\begin{array}{lll}
  - \epsilon_{1,2}, &  z < z_0  \\
    \frac{\epsilon_{1,2} }{2}
    \left[ \cos \left( \pi \frac{z_c - z}{z_c-z_0} \right)-1 \right],
     &   z_0 \le z < z_c
\end{array}
\right.
\end{equation}
where an excellent agreement is
obtained for $\epsilon_{r} = 1.5\, \rm eV$, $\epsilon_{0}= 0.7\, \rm
eV$, $\epsilon_{1} = -7/45\, \rm eV$, and $\epsilon_{2} = -2/45\, \rm
eV$. The $U_i(z)$ prefactors, as well as the final potential and the {\em
  ab~initio} data, are shown in
Fig.~\ref{fig:fitcpmd}(a).
%
%
The contour plot of the surface potential is depicted in
Fig.~\ref{fig:fitcpmd}(b).
%

\section{Simulation details}

BPA-PC melt is confined to a slit pore comprised of two parallel (111)
nickel surfaces. The pore width is set to $L_{z} = 48.54 \sigma$. The
normal to the surfaces is parallel to the $z$ axis. The sample
consists of $240$ BPA-PC molecules of $N=20$ repeat units.  Because of
the small entanglement length of BPA-PC ($N_e = 6$) these chains are
already weakly entangled. The units are chosen such that $kT =1$ with
$T=570\, K$, i.~e. $1\, {\rm eV}$ corresponds to about $20\,kT$; the
unit of length is $\sigma = 4.41\, {\rm \AA}$.
%
Periodic boundary conditions are employed in $x$ and $y$ directions.
The $x$ and $y$ box dimensions are set to $L_{x} = 22.23\, \sigma$ and
$L_{y} = 21.72\, \sigma$, which corresponds to a (111) hexagonal
lattice of nickel with $39$ and $22$ unit cells (the unit lattice
constant of the hexagonal lattice is about $a = 0.57\, \sigma$).  The
system density is fixed at $0.85 \sigma^{-3}$, which corresponds to
$1.05\, {\rm g/cc}$, the experimental density at $T=570K$.
Starting configurations are generated by randomly placing the chains
in the simulation box. A short run is then used to remove the
bead-bead overlaps~\cite{abrams2003b}.  Then the system is equilibrated
during $10^8$ timesteps. The production run is performed in the $NVT$
ensemble with Langevin thermostat with friction $0.5
\tau^{-1}$. The thermostat is switched off in the
shear direction (along the $x$ axis).  The velocity-Verlet algorithm
with the timestep $\Delta t = 0.005 \tau$ is used to integrate the
equations of motion.
The top and bottom walls move (in opposite directions) at constant
velocity $v_w$, so that the shear rate is $\dot{\gamma} =
2v_{w}/L_z$.
%
%
We use
$v_{w} \tau/\sigma = 0$, $0.001$, $0.01$, and $0.1$ or $s =
\dot{\gamma} \tau_{r} \approx 0$, $2$, $20$, and $200$, respectively,
where ${\tau}_{r} \approx 5 \times 10^4 {\tau}$ is the characteristic
relaxation time of a single chain of BPA-PC in bulk~\cite{leon2005}.
Note that we apply rather high shear rates, of the order of $10^5\, \rm s^{-1}$. 
However, the average chain length in a BPA-PC melt is
$N \approx 70$ and, therefore, the corresponding chain reptation time $\tau_d \sim
N^{3.4}$ is almost by two orders of magnitude bigger than that of the
$N=20$ chains.  Equivalently, shear rates would be reduced by almost two orders 
of magnitude, close to the values used to process the melt.

\section{Results}

\begin{figure}
\onefigure[height=3.8cm]{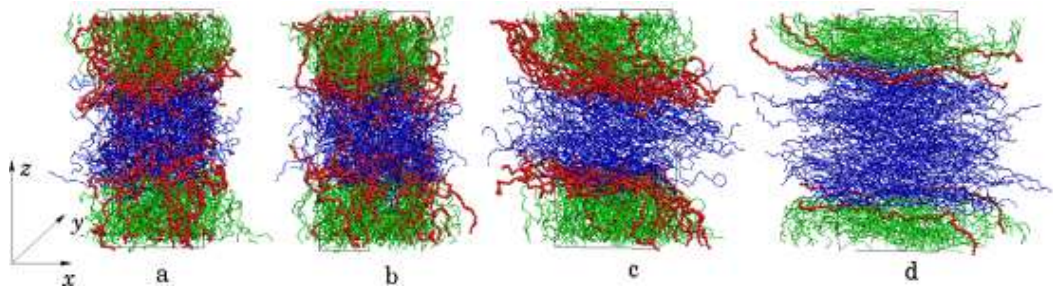}
\caption{
  Snapshots of the system for different wall velocities: (a) no shear;
  (b) $v_w = 0.001 \sigma / \tau$; (c) $v_w = 0.01 \sigma / \tau$; (d)
  $v_w = 0.1 \sigma / \tau$. Polymer chains are divided on three
  populations: chains which adsorb both ends (green), only one end
  (red), and no ends on the surface (blue). }
\label{fig:snapshot}
\end{figure}

The snapshots of the systems, for different shear rates, are shown in
Fig.~\ref{fig:snapshot}. 
%
The increase of the shear rate results in chain
stretching. For small shear rates,
Fig.~\ref{fig:snapshot}(b), the change in the chain configurations is
practically negligible compared to the system without shear,
Fig.~\ref{fig:snapshot}(a). For medium shear rates,
Fig.~\ref{fig:snapshot}(c), the chain conformations, as well as the
width of the layer formed by the two-end adsorbed chains, does not
change.  The chains adsorbed with one end stretch and form a thin
layer between the bulk and the chains adsorbed with two ends. The
chains in the bulk decouple from this layer.  Finally, for even higher
shears, Fig.~\ref{fig:snapshot}(d), the single-end attached chains
stretch even more; almost all of them desorb and move into the
bulk. As a result, the layer made of adsorbed chains shrinks.

Keeping this qualitative picture in mind, let us quantify our
results. We first have a look at the changes in the bead density
profiles, shown in Fig.~\ref{fig:density}. For medium shear rates
($v_{w} = 0.01\, \sigma / \tau$, $ \dot{\gamma} \tau_{\rm
r}=20$) the layering of the beads next to the wall extends to
a depth of about the radius of gyration $R_g \sim 7 \sigma$,
similar to the system without shear. The chain end density has a
large peak next to the wall; then the region with practically no
ends (depleted region) follows; finally, the bulk concentration of
the ends is reached at a distance of about $2R_g$.
\begin{figure}
\twofigures[height=3.8cm]{density.eps}{density_com.eps}
%
%
\caption{
  Number density profiles of beads in a sheared melt for $s =
  \dot{\gamma} \tau_{r} = 20$.  Insets zooms in the chain-end
  density profiles at small and large distances from the wall, for
  different shear rates.  }
\label{fig:density}
\caption{
  Center of mass density profiles, showing division into populations
  of chains which adsorb both ends to the surface (green), one
  end on the surface (red), and the total center of mass density
  (black).  Insets illustrate changes in the profiles for different
  shear rates.}
\label{fig:density_com}
\end{figure}
If we increase the shear rate, the density of the ends at the wall
decreases, as shown in the left inset of Fig~\ref{fig:density}. At the
same time, the depleted layer shrinks, as illustrated in the right
inset of the same figure. Both effects are due to desorption of the
chain ends from the walls, as well as stretching and tilting of the
chains under shear.

More information about the chain conformations at the surface and in
the bulk can be obtained from the center of mass density profiles,
shown in Fig.~\ref{fig:density_com}. The chains are divided into two
categories: those chains with two ends adsorbed and those chains with
only one end adsorbed on the wall.
Without shear, the profiles are similar to those reported
before~\cite{abrams2003b}: any chain with the center of mass in the
layer next to the wall has both ends adsorbed on the surface. The next
layer contains a majority of chains with a single end adsorbed. The
single-end adsorbed chains mix well with the bulk of the melt.
For small shear rates ($\dot{\gamma}\tau_{r} = 2$) the
changes in the layers are not detectable. However, for higher shear
rates, several different scenarios develop.
For $\dot{\gamma}\tau_{r} = 20$, the layer of
single-end adsorbed chains becomes more narrow; the
center of mass density drops down almost to zero immediately after it,
i.~e. a layer of single-end adsorbed chains is formed on the top of
the layer of double-end adsorbed chains; the bulk chains detach from
these two layers. Again, the layer of the adsorbed (both with one and
two ends) chains shrinks compared to the static case, due to
stretching and alignment of adsorbed chains.
If we increase the shear rate ($\dot{\gamma}\tau_{r} =
200$), the single-end adsorbed chains desorb and
%
%
additional layers due to the chain dense packing at the wall
appear.

\begin{figure}
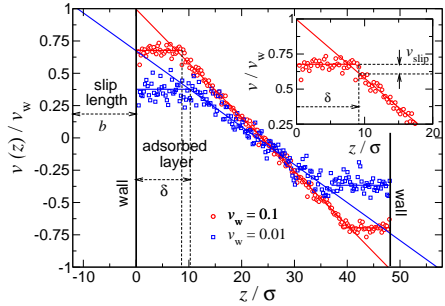

\onefigure[height=4cm]{velocity.eps}
\caption{
  The normalized velocity profiles for two shear rates. Symbols:
  simulation results.  Solid lines: fit to a constant or a linear
  function, depending on the layer.}
\label{fig:velocity}
\end{figure}

Finally, Fig.~\ref{fig:velocity} shows the velocity profiles
normalized to the wall velocity. For all shear rates, the profiles
have similar features:
(i) next to the wall it is practically constant, i.~e. the adsorbed
chains are dragged by the wall. The flat plateau has the same width as
the adsorbed layer, measured from the center of mass density profiles,
Fig.~\ref{fig:density_com}. Immediately after the plateau the velocity
profile becomes a linear function of $z$. The transition region is
very small, again due to the fact that the adsorbed and bulk polymer
chains are very much decoupled.
(ii) the velocity of the beads at the walls ($v_s$) is smaller
than the wall velocity, i.~e. even though the chain ends strongly
adsorb on the wall they can still slide over it. This is not
surprising - an adsorbed bead can hop between the hollow to bridge
sites, since the difference in the adsorption energies of
these two sites is small, of the order of $0.1\, {\rm eV}$.

Let us denote the thickness of the adsorbed layer as $\delta$ and
the velocity of this layer as $v_s$. Then the bulk velocity can be
written as $v = v_s (L-2z)/(L-2\delta)$. The slip length $b$,
obtained from the condition $v(z = -L/2 - b) = v_w$, reads
\begin{equation}
b = \left( v_w / v_s -1 \right) L/2 - \delta v_w / v_s.
\label{eq:slip}
\end{equation}
It can be seen that two mechanisms contribute to the total slip length
$b$. The first one is due to the apparent slip of the adsorbed layer
over the surface. The second, negative, contribution is due to the
finite thickness of the adsorbed polymer layer. Examining the velocity
profiles we see that for $s=20$ the apparent slip dominates and
results in a slip length of the order of $10 \sigma$. For higher shear
rates, $s=200$, the two contributions compensate each other, even
though $\delta$ decreases, again due to the fact that $v_s/v_w$
increases with the increase of the shear rate.
Another possible contribution to the total slip length is shown in the
inset of Fig.~\ref{fig:velocity}. Here the bulk of the melt slips over
the wall-adsorbed layer. This contribution is, however, rather small.
Finally, for the friction coefficient $\beta = f_x /(v_w-v_s) $,
where $f_x$ is the friction force per unit areal, we obtain $\beta (s = 20) = 4.4$
and $\beta (s = 200) = 3.9$ ($\beta$ is given in units of $kT\, \tau
\, \sigma^{-4}$). Therefore, in spite of the fact that the higher
shear rates lead to a reduction of slip length, the surface friction
itself decreases. In fact, it is the disentanglement of the bulk
polymer chains from the surface layer which is responsible for this
decrease. 

To conclude, we have studied a shear of a BPA-PC melt over a (111) nickel surface as a specific test case of a more general phenomenon of shear of an adsorbing polymer melt over structured surfaces. We find that two mechanisms contribute to the kinetic friction and effective hydrodynamic boundary conditions. 
The first mechanism is similar to the one observed when an adsorbed surface layer (either solid or liquid) slides over a structured substrate~\cite{cieplak1994}. The substrate potential induces the density modulation in this layer, and energy is lost from these modulations through the anharmonic coupling to the thermostat. In our case the attached chain ends move along surface minimal activation paths and the energy loss is due to their scattering on the wall potential and coupling of the rest of the chain to the thermostat.
On the other hand, we also observe that the one end-attached chains undergo a coil/stretch transition and disentangle from the melt, in agreement with the theoretical predictions for grafted chains~\cite{brochard1992,ajdari1994}. In principle, stretching should lead to large slip lengths at high shear rates. For our system, however, the net effect of chain stretching is much smaller than the energy dissipation in the adsorbed layer.  
All together, competition between these two mechanisms, combined with the ability of chains to desorb, results in a nontrivial friction law, with both the slip length and the friction force depending on the shear rate. 

Possible implication of our work
is that the surface-anchored layers can be efficiently used to adjust
the friction between polymer melt and a surface. This can be done by
changing the surface concentration of one- and two-end adsorbed chains
by, for example, additives occupying the attractive sites of the
nickel surface~\cite{andrienko2005}.


\acknowledgments

This work was supported by the Alexander von Humboldt Foundation,
Germany (X.Z.) and by the BMBF, under Grant No. 03N6015. Advise of
B.~D{\"u}nweg is acknowledged.


\begin{thebibliography}{17}

\bibitem{klein1996}
  \Name{J.~Klein}
  \REVIEW{Annu. Rev. Mater. Sci.}{26}{1996}{581}.

\bibitem{jin1993}
  \Name{Z.~M.~Jin \etal}
  \REVIEW{Wear} {170} {1993} {281};
%
  \Name{X.~Ma}
  \REVIEW{IEEE Trans. Magn} {35}{1999}{2454}.

\bibitem{grest1999}
  \Name{G.~S.~Grest}
  \REVIEW{Adv. Polym. Sci.}{138}{1999}{149};
%
  \Name{B.~N.~J. Persson}
  \REVIEW{Surf. Sci. Rep.}{33}{1999}{85};
%
  \Name{J. Ringlein \and M. O. Robbins}
  \REVIEW{Am. Journ. Phys.}{72}{2004}{884}.

\bibitem{yamamoto2004}
  \Name{R.~Yamamoto \and A.~Onuki}
  \REVIEW{Phys. Rev. E} {70}{2004}{041801}.

\bibitem{everaers2004}
  \Name{R.~Everaers \etal}
  \REVIEW{Science} {303}{2004}{823}.

\bibitem{leger1999}
  \Name{L.~L{\'e}ger, E.~Rapha{\"e}l, \and H.~Hervet}
  \REVIEW{Adv. Polymer Sci.} {138} {1999} {185}.


\bibitem{guiselin1992}
  \Name{M. Aubouy, O. Guiselin \and E. Raphael}
  \REVIEW{Macromolecules}{29}{1996}{7261}.

\bibitem{smith2005}
  \Name{K. A. Smith, M. Vladkov \and J.-L. Barrat}
  \REVIEW{Macromolecules}{38}{2005}{571}.

\bibitem{ajdari1994}
  \Name{A. Ajdari \etal 
  }
  \REVIEW{Physica A}{204}{1994}{17}.

\bibitem{brochard1992}
  \Name{F. Brochard \and P. G. Degennes}
  \REVIEW{Langmuir}{8}{1992}{3033}.  

\bibitem{barrat1999}
  \Name{J. L. Barrat \and L. Bocquet}
  \REVIEW{Phys. Rev. Lett.}{82}{1999}{4671}.  

\bibitem{thompson1997}
   \Name{P. A. Thompson, \and S. M. Troian}
   \REVIEW{Nature}{389}{1997}{360}.
    
\bibitem{cottinbizonne2003}
  \Name{C. Cottin-Bizonne \etal} 
  \REVIEW{Nature Materials}{2}{2003}{237}.     

\bibitem{brochardwyart1994}
  \Name{F. Brochard-Wyart \etal} 
  \REVIEW{Langmuir}{10}{1994}{1566}.
  
\bibitem{abrams2003}
  \Name{C.~F.~Abrams \and K.~Kremer}
  \REVIEW{Macromolecules} {36} {2003} {260}.

\bibitem{dellesite2002}
  \Name{L.~Delle~Site, C.~F.~Abrams, A.~Alavi, \and K.~Kremer}
  \REVIEW{Phys. Rev. Lett.} {89} {2002} {156103}.

\bibitem{abrams2003b}
  \Name{C.~F. Abrams, L.~Delle~Site, \and K.~Kremer}
  \REVIEW{Phys. Rev. E} {67} {2003}{021807}.

\bibitem{dellesite2004}
  \Name{L.~Delle~Site, S.~Leon, \and K.~Kremer}
  \REVIEW{JACS} {126} {2004} {2944}.

\bibitem{tschop1998}
  \Name{W.~Tsch{\"o}p, K.~Kremer, \etal}
  \REVIEW{Acta Polym.} {49} {1998}{61}.

\bibitem{dellesite2003}
  \Name{L.~Delle~Site, A.~Alavi, \and C.~F.~Abrams}
  \REVIEW{Phys. Rev. B} {67} {2003} {193406}.


\bibitem{leon2005}
  \Name{S.~Leon, L.~Delle~Site, \and K.~Kremer}
  \REVIEW{in preparation}{}{2005}.

\bibitem{cieplak1994}
  \Name{M.~Cieplak, E.~D.~Smith \and M.~O.~Robbins}
  \REVIEW{Science}{265}{1994}{1209}.

\bibitem{andrienko2005}
  \Name{D.~Andrienko,  S.~Leon,  L.~Delle~Site, \and  K.~Kremer}
  \REVIEW{in preparation}{}{2005}.
 

\end{thebibliography}

\end{document}